\documentclass[a4]{aa}
\usepackage{longtable,lscape}
\usepackage{graphicx,amssymb,latexsym,amsmath}
\usepackage[authoryear]{natbib}
\bibpunct{(}{)}{;}{a}{}{,} 
\newcommand{\gtsim}{\protect\raisebox{-0.5ex}{$\:\stackrel{\textstyle >}
        {\sim}\:$}}

\newcommand{\kms}{\,km\,s$^{-1}$}

\usepackage[authoryear]{natbib}
\bibpunct{(}{)}{;}{a}{}{,} 
 
\begin{document}

\title{Photospheric and chromospheric activity on the young solar-type star HD\,171488 (\object{V889~Herculis})\thanks{Based 
on observations collected at Calar Alto Astronomical Observatory (Spain) and Catania Astrophysical Observatory (Italy).}
\fnmsep\thanks{Table \ref{tab:phot} and Figure \ref{fig:spe} are only available in electronic form at {\tt http://www.aanda.org}}}
   
\author{A. Frasca\inst{1} \and K. Biazzo\inst{1,2} \and Zs.~K\H{o}v\'{a}ri\inst{3} \and E. Marilli\inst{1} \and 
\"{O}. \c{C}ak{\i}rl{\i}\inst{4,5}}
\offprints{A. Frasca}
\mail{antonio.frasca@oact.inaf.it}

  \institute{INAF, Osservatorio Astrofisico di Catania, via S. Sofia, 78, I-95123 Catania, Italy
\and INAF, Osservatorio Astrofisico di Arcetri, L.go E. Fermi, 5, I-50125 Firenze, Italy
\and Konkoly Observatory, H-1525 Budapest, P.O.Box 67, Hungary
\and Ege University, Science Faculty, Astronomy and Space Sciences Dept., 35100 Bornova, \.{I}zmir, Turkey
\and T{\"U}B{\.I}TAK National Observatory, Akdeniz University Campus, 07058 Antalya, Turkey}

   \date{Received --- / Accepted --- }

\abstract
{}
{We present the results of contemporaneous spectroscopic and photometric monitoring of the young solar-type
star \object{HD~171488} ($P_{\rm rot}\simeq 1.337$ days) aimed at studying surface inhomogeneities at both 
photospheric and chromospheric levels.}
{Echelle FOCES spectra ($R\simeq$\,40\,000) and Johnson $BV$ photometry have
been performed in August 2006, with a good coverage of rotational phases. Spectral type, rotational velocity, 
metallicity, and gravity were determined using a code developed by us (\textsc{Rotfit}) and a library of spectra at the same resolution of
slowly-rotating reference stars.  The metallicity was measured from the analysis of iron lines with the \textsc{Moog} code.
The spectral subtraction technique was applied to the most relevant chromospheric diagnostics included in the FOCES spectral range,
namely \ion{Ca}{ii} IRT, H$\alpha$, \ion{He}{i}\,D$_3$, H$\beta$, and \ion{Ca}{ii} H\&K lines. }
{A simple model with two large high-latitude spots is sufficient to reproduce the $B$ and $V$ light curves 
as well the radial velocity modulation, if a temperature difference between photosphere and spots of about 1500\,K is used. A Doppler imaging analysis 
of photospheric lines basically confirms a similar spot distribution. 
With the help of an analogous geometric two-spot model, we are able to reproduce the observed modulations in the residual 
chromospheric emissions adopting different values of ratios between the flux of plages and quiet chromosphere (about 5 for H$\alpha$ and 3 for 
\ion{Ca}{ii} diagnostics).
Facular regions of solar type appear to be the main responsible for the modulations of chromospheric diagnostics. 
Both the spot/plage model and the cross-correlation between the light curve and the chromospheric line fluxes display a sensible lead effect of plages 
with respect to spots (from 20$\degr$ to 40$\degr$ in longitude), as already observed in some active solar-type stars and RS CVn systems.}
{The contemporaneous monitoring of photospheric and chromospheric diagnostics in the young and rapidly rotating solar-type star HD~171488
allowed us to detect active regions which have nearly the same location at both atmospheric layers, with plages slightly
leading spots in longitudes. These active regions are similar to the solar ones in some respect, because the spot temperature is close to that
of sunspot umbrae and the plage flux-contrast is consistent with the average solar values. The main differences with respect to the Sun are 
larger sizes and higher latitudes. }

\titlerunning{Photospheric and chromospheric activity on HD\,171488}
\authorrunning{A. Frasca et al.}
\maketitle

\section{Introduction}

\label{sec:Intro}

Young solar-mass stars just arrived on the Zero-Age Main Sequence (ZAMS) or on their way 
to it are in a very important phase of their life. Indeed, at this time they 
start to spin up approaching the ZAMS when they get free from their circumstellar disks. 
Thereafter, the disks can start to ``condense" giving rise to proto-planetary systems 
and the stars experience angular momentum loss resulting from a magnetized stellar wind. 
It is therefore very important to define the physical conditions that 
affect the subsequent evolution of the star and of its environment. 
In addition to the basic stellar parameters ($T_{\rm eff}$, $\log g$, [Fe/H]), it is of fundamental 
importance to know the rotation rate and the level and behaviour of the star's magnetic activity. 

Fast rotators with age of about 100 Myr are likely originated by a less efficient disk locking during
their T~Tau phase, but they also suggest  a reduced effect of magnetic braking. 
It has been proposed that the rate of angular 
momentum loss decreases or, at least, saturates at high rotation rates (\citealt{MacGregor91}; 
\citealt{Barnes96}). Saturation of magnetic activity is witnessed by the typical behaviour of 
X-ray emission and other activity signatures as a function of the rotation speed (\citealt{Randich97}; 
\citealt{Krishna97}). Periodic variability of saturated fast-rotating stars is present both in 
the optical continuum 
(e.g., \citealt{Marilli97}; \citealt{Stassun04}) and in the X-ray band (\citealt{Flaccomio05}). This rises 
questions about the true extent of saturation in fast-rotating stars. 
It is intriguing that very fast rotating stars, for which a ``super saturation" regime, i.e. a slight decline of 
magnetic activity with the increase of rotation rate, is displayed  \citep{Prosser96}, still show organized magnetic 
topologies but inefficient angular momentum losses \citep{Stepien01}.

For all these reasons we consider it mandatory to investigate the surface topology of magnetic active 
regions on fast-rotating young stars, also because of the implications on the stability and early 
evolution of planetary systems.

HD~171488 (=\object{V889~Her}) is a young fast-rotating solar analogue that has been the subject of several 
studies in the last decade. Its kinematic properties indicate that HD~171488 is a member of the Local Association,
a stream of young stars with ages ranging from 20 to 150 Myr \citep{Montes01}.
It is probably the brightest single early-G type star that is rotating fast enough 
for mapping its photosphere through Doppler Imaging (DI).
\citet{Strass03} made the first dedicated photometric and spectroscopic study of HD~171488 in which they defined 
its astrophysical parameters, concluding that it is a single G0V main-sequence star with an age of about 30--50
Myr based on  both its position on the HR diagram and the lithium content. They also made the first reconstruction
of its photosphere through the DI technique, showing a big polar spot with additional high-latitude features.
Subsequent works based on DI have always found a polar spot \citep{Marsden06,Jeffers08}. From long-term photometry
and DI, two active longitudes separated by about 180$^{\circ}$ seem to be persistent for several years 
\citep{Jarvi08,Huber09}.
\citet{Jeffers08} found a high solar-type differential rotation with the equator lapping the poles every 12-13 days.
\citet{Jarvi08} found instead a much weaker differential rotation.

Despite the number of works devoted to this object in the last few years, which are mainly based on DI or Zeeman-DI,
a simultaneous detailed investigation of the chromospheric and photospheric inhomogeneities has not been performed
so far.
 
In the present work we analyze contemporaneous $BV$ photometry and high-resolution spectra of HD~171488.
We determined the main parameters of the photospheric spots by means of the DI technique applied
to a few suitable absorption lines and of a spot model for simultaneous solution of light and radial velocity curves.
The behaviour of chromospheric inhomogeneities was investigated thanks to the Ca\,{\sc ii} H\,\&\,K ($\lambda=$3968.49 
\AA, 3933.68 \AA), H$\epsilon$ ($\lambda=$ 3970.074 \AA), H$\alpha$ ($\lambda=$6562.849 \AA), and
Ca\,{\sc ii} infrared triplet (IRT; $\lambda=$8498.06 \AA, 8542.14 \AA, 8662.17 \AA) lines.

The work is organized as follows. In Section\,\ref{sec:Obs} we describe the observations and data reduction. 
The spectral classification, rotational and radial velocities, evolutionary status, and metallicity are discussed in 
Section\,\ref{sec:AP}.
The diagnostics of photospheric and chromospheric activity are analyzed in Sections \ref{sec:phot_act} and
\ref{sec:chrom_act}, respectively.  A short discussion on the behaviour of photospheric and chromospheric activity is
presented in Section\,\ref{sec:discussion}.
Section \ref{sec:concl} contains our conclusions.

\section{Observations and reduction}

\label{sec:Obs}

\subsection{Spectroscopy}
We observed HD~171488  at the 2.2-m Cassegrain telescope of the Calar Alto Observatory (CAHA, Sierra de Los Filabres, Spain) 
with the Fiber Optics Cassegrain {\it \'Echelle} Spectrograph (FOCES; \citealt{Pfeiffer1998}) during four nights from 13 
to 16 August, 2006. The 2048$\times$2048 CCD detector Site\#1d (pixel size = 24\,$\mu$m)	
allowed us to achieve, with an exposure time of 15 minutes, a signal-to-noise ($S/N$)
ratio in the range 80--160 for HD~171488 ($V\simeq7^{\rm m}.4$), depending on airmass and sky conditions. 

The spectral resolution, as evaluated from the full width at half maximum (FWHM) of the emission lines of the 
Th-Ar calibration lamp, was in the range 0.15--0.22\,\AA\ from the blue to the red, yielding a 
resolving power $R=\lambda/\Delta\lambda\,\simeq\,$40\,000.

The data reduction was performed by using the {\sc echelle} task of the IRAF\footnote{IRAF is distributed by the 
National Optical Astronomy Observatory, which is operated by the Association of the Universities for Research in 
Astronomy, inc. (AURA) under cooperative agreement with the National Science Foundation.} package, following the
scheme summarized by \citet{Biazzo09}.

To remove the telluric water vapor lines at the H$\alpha$ and Na\,{\sc i}\,D$_2$ wavelengths, we used the procedure
described by \citet{Frasca00}, adopting as telluric template a high-S/N spectrum of \object{$\alpha$~Cep} 
(A7~IV, $v\sin i=196$~km\,s$^{-1}$), acquired during our observing run.  

\subsection{Photometry}
The photometric observations were performed in the $B$ and $V$ Johnson filters with the 91-cm 
Cassegrain telescope at the {\it M. G. Fracastoro} station ({\it Serra La Nave}, Mt. Etna, Italy) of the 
{\it Osservatorio Astrofisico di Catania} (OACt). The observations were made with a photon-counting 
refrigerated photometer equipped with an EMI 9893QA/350 photomultiplier, cooled to $-15\,^\circ$C. 
The dark current noise of the detector, operated at this temperature, is about $1$ count/sec.

HD~171488 was observed from 14 to 21 August 2006 for a total of 8 nights, along with \object{HD~171286} ($V=6^{\rm m}.84$, 
$B-V=1.06$, \citealt{Oja87}) used as a comparison star ($C$) for the differential photometry and \object{HD~171623} ($V=5^{\rm m}.79$, 
$B-V=0.00$, present work) and \object{BD+17~3634} ($V=8^{\rm m}.43$, $B-V=0.57$, present work) as check ($Ck_1$, $Ck_2$) stars.
We adopted HD~171286 as the local standard for the determination of the photometric instrumental ``zero points''. 
Several standard stars, selected from the list of \citet{Lan92}, 
were also observed during the run in order to determine the transformation coefficients 
to the Johnson standard system. The observed magnitudes were corrected for atmospheric extinction using the seasonal 
average coefficients for the Serra La Nave Observatory. 

The data were reduced by means of the photometric data reduction package {\sc phot} designed for 
the photoelectric photometry of the OACt \citep{LoPr93}. The photometric errors, estimated from 
measurements of standard stars with a brightness comparable to the program stars, are typically 
$\sigma_V \approx 0^{\rm m}.010$ and $\sigma_{B-V} \approx 0^{\rm m}.014$.

In order to improve the photometric precision we averaged four consecutive measurements of the variable 
and adopted the standard deviation, ranging from about 0.005 to 0.015 mag, as an error estimate for each of 
the mean photometric points.

The photometric data are reported in Table~\ref{tab:phot}\footnote{Available as on-line material.}.

\section{Astrophysical parameters}
\label{sec:AP}

Thanks to the high resolution and the wide wavelength coverage of the FOCES spectra, we have redetermined spectral type, 
effective temperature, gravity, metallicity, radial and rotational velocity and quoted these values in 
Table~\ref{tab:literature_param} along with literature values. The evolutionary status of HD~171488 has also been checked. 

\begin{table}
\caption{Physical parameters of HD~171488 from the literature and present work.}
\centering
 \begin{tabular}{lll}
  \hline\hline
  \noalign{\smallskip}
  Parameter                  & Value & Reference\\
  \noalign{\smallskip}
  \hline
  \noalign{\smallskip}
  Spectral Type                   & G0V   & \cite{Harlan69}\\
   "                              & G2V    & \cite{Cuti02}\\
   "                              & G2V    & Present work\\
  $v\sin i$ (km\,s$^{-1}$)        & 38  	   & \cite{Fekel97}\\
  "                               & 39.0$\pm$0.5   & \cite{Strass03}\\
  "                               & 37.5$\pm$0.5   & \cite{Marsden06}\\
  "                               & 38.0$\pm$0.5   & \cite{Jeffers08}\\
  "                               & 37.1$\pm$1.0   & Present work \\
  $P_{\rm rot}$ (days)            & 1.3371$\pm$0.0002  & \cite{Strass03}\\  
  "                               & 1.313$\pm$0.004  & \cite{Marsden06}\\
  "                               & 1.33697          & \cite{Jarvi08}\\
  $W_{\textrm{Li\,{\sc i}}}$ (m\AA)   & 220                & \cite{Cuti02}\\
  "                             & 213$\pm$7	     & \cite{Strass03}\\
  "                             & 215$\pm$7	     & Present work \\
  $T_{\rm eff}$ (K)               & $5830\pm50$  & \cite{Strass03}\\
  "                               & $5808$   & \cite{Jeffers08}\\
  "                               & $5750\pm130$ & Present work (\textsc{Rotfit}) \\
  "                               & 5850$\pm$100 & Present work (\textsc{Moog}) \\
  $\log g$                        & 4.50  & \cite{Strass03}\\
  "                               & $4.30\pm0.15$ & Present work\\  
  ${\rm [Fe/H]}$			  & $-0.5$  & \cite{Strass03}\\
  "				  &  +0.01$\pm$0.09  & Present work (\textsc{Rotfit}) \\
  "				  &  +0.15$\pm$0.07  & Present work (\textsc{Moog}) \\
  \noalign{\smallskip}
  \hline
\end{tabular}
\label{tab:literature_param}
\end{table}

\subsection{Spectral type and rotational velocity}
\label{sec:spec_type}

The spectral type and the $v\sin i$ of HD~171488 was derived through the IDL\footnote{IDL (Interactive Data Language) is a 
registered trademark of ITT Visual Information Solutions.} code \textsc{Rotfit} \citep{Frasca06}. 
The code simultaneously finds the spectral type and the $v\sin i$ of the 
target searching for, into a library of standard star spectra, the spectrum which is best fitting 
(minimum of the residuals) the target one, after the rotational broadening by convolution with a 
rotational profile of increasing $v\sin i$ at steps of 0.5\,km\,s$^{-1}$. 
We acquired spectra of a very small sample of standard stars (9 objects) with FOCES during our run.
Thus, we preferred to use a library of 185 ELODIE Archive standard stars well distributed in effective temperature, 
spectral type, and gravity, and in a suitable range of metalicities (\citealt{Prugniel01}) that have nearly the
same resolution ($R\simeq$\,42\,000) as our FOCES spectra. 
We found for HD~171448 a G2V spectral type, a nearly solar metallicity, [Fe/H]=+0.01$\pm$0.09, $\log g=4.30\pm0.15$, 
and $T_{\rm eff}=5750\pm130$\,K (see Table~\ref{tab:literature_param}), as average parameters of the best ten 
standard stars per each echelle order. This temperature range encompasses all the previous $T_{\rm eff}$ values from the 
literature.

We remark that our temperature determination is not strongly affected by starspots, due to their low contribution to the observed
flux in the spectral range of the observations. Indeed, as shown by \citet{Frasca05}, for large spot--photosphere temperature 
differentials the star's integrated emission is dominated by non-spotted photospheric flux.
In their Appendix A, \citet{Frasca05}  outline a formalism for evaluating the mean temperature observed from a star in
a given spectral region with a given spot filling factor and photosphere--spot temperature difference, $\Delta T$. 
Adopting $T_{\rm eff}=5800$\,K, $\Delta T=1500$\,K, and a filling factor of 0.1, that is likely an overestimation, 
the measured temperature would be, at most, only 50\,K cooler than the true photospheric temperature.
Thus we conclude that the ``undisturbed'' photospheric temperature must be of about 5800\,K and we adopted this 
value for the following analysis. 

For the $v\sin i$ determination, we chose as ``non-rotating'' templates three stars with spectral type
similar to HD~171488 observed with 
FOCES during the same run, namely \object{54~Aql} (F8V), \object{10~Tau} (F9V-IV), and \object{72~Her} (G2V).
We found, as the average of different echelle orders and these three templates, $v\sin i=37.1\pm$1.0\,\kms, which is 
in very good agreement with the literature values reported in Table~\ref{tab:literature_param}. 
However, using the \textsc{Rotfit} code with the ELODIE templates, we found a nearly equal value of 37.4\,\kms.  

Examples of the application of the \textsc{Rotfit} code to two different spectral regions are shown in 
Fig.~\ref{fig:spe}\footnote{Available in electronic form only.}, where the very good agreement between observed 
and standard spectra is apparent.

\subsection{Radial Velocity}
\label{sec:radvel}

We measured the heliocentric radial velocity ($V_{\rm r}$) by means of the cross-correlation technique, 
(e.g., \citealt{Simkin1974}, \citealt{Gunn1996}), taking advantage of the wide spectral coverage offered by FOCES. 
The radial velocity standard star $\alpha$ Ari ($V_{\rm r}=-14.2$\,\kms), observed in the same run, was 
used as template.  
We cross-correlated each spectral order of the FOCES spectra of HD~171488 with the template by using the IRAF task {\sc fxcor}, 
avoiding the orders with low $S/N$ ratio or contaminated by broad and/or chromospheric lines (e.g., H$\alpha$, Na\,{\sc ii}\,D$_2$, 
Ca\,{\sc ii} H\&K) or by prominent telluric features. 
We ended up with 60 orders useful for the calculation of the cross-correlation functions (CCFs). 
The radial velocities listed in Table~\ref{tab:eqw_lines} per each spectrum are weighted averages of the values of the 60 chosen 
orders with weights $w_i = 1/\sigma_i^2$, where $\sigma_i$ is the error for the $i$-th order evaluated by {\sc fxcor}.
The average value over the entire observing run is $V_{\rm r}=-23.3\pm0.2$ km\,s$^{-1}$, which is in 
close agreement with previous determinations.	
Anyway, the individual values of $V_{\rm r}$ display a clear rotational modulation (see Fig.~\ref{fig:modulations}) as already found by 
\citet{Huber09} and explained in terms of a ``Rossiter-McLaughlin'' effect caused by starspots instead of eclipses, as in the original
formulation (e.g., \citealt{Rossiter1924}).

\subsection{Evolutionary status, lithium abundance and age}
\label{sec:HR_diagram}

\begin{figure}
\centering
\includegraphics[width=8cm]{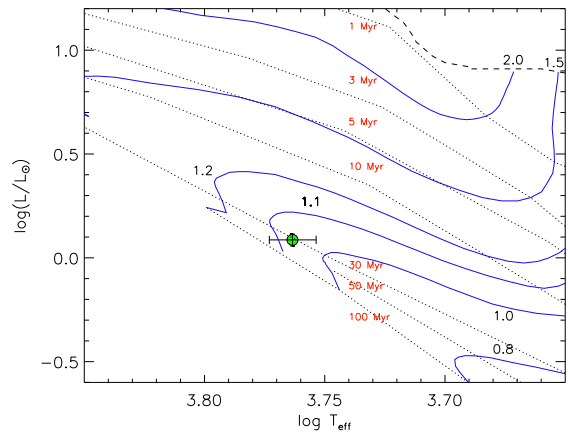}
\caption{Position of HD~171488 on the HR diagram. The evolutionary tracks and isochrones of \citet{palla1999} 
are shown by continuous and dotted lines, respectively. The birth line of \citet{palla1999} is displayed by a 
dashed line.} 
\label{fig:hr_diagram}
\end{figure}

We checked the evolutionary status of HD~171488 by means of its position onto the Hertzsprung-Russell (HR) 
diagram (Fig.~\ref{fig:hr_diagram}) and the lithium content. We used the evolutionary tracks calculated by \citet{palla1999} for the pre-main 
sequence phase.
The unspotted photospheric temperature of 5800\,K was adopted.
Firstly, to derive the stellar luminosity, we evaluated the interstellar extinction $A_{\rm V}\simeq 0.03$\,mag (quite negligible) from 
the Hipparcos parallax ($\pi=26.87\pm0.89$\,mas, \citealt{HIPPA97}), assuming a mean extinction of 0.7 mag/kpc.
 Then, the de-reddened $V_0$ magnitude at maximum brightness (spottedness minimum) 
was converted into absolute magnitude $M_{\rm V}$ with the parallax and subsequently converted into bolometric
magnitude by using the bolometric correction tabulated by \citet{Flower96} as a function of 
the effective temperature. The bolometric magnitude of the Sun, $M_{\rm bol} = 4.64$ \citep{Cox00}, was used to
express the stellar luminosity in solar units ($L\simeq 1.2L_{\odot}$).

The comparison with the evolutionary tracks and isochrones of \citet{palla1999} indicates that HD~171488 is a post-T~Tau or 
ZAMS star with an age of $\approx$\,50 Myr, a mass of about 1.08\,$M_{\odot}$, and a radius $R\simeq\,1.1\,R_{\odot}$, in good agreement 
with previous determinations. The comparison with the \citet{Siess00} evolutionary tracks leads to slightly larger values of  mass 
($M\simeq 1.16\,M_{\odot}$) and  age (70--100 Myr) that are, however, still consistent with the former within the errors.  

A further evidence of the youth of our object is given by the presence of the \ion{Li}{i} $\lambda6707.8$ absorption line.
Indeed, lithium is strongly depleted from the stellar atmospheres of late-type stars when mixing mechanisms pull
it deeply in their convective layers and a deep Li\,{\sc i}\,$\lambda$6707.8 photospheric 
absorption line is generally considered as a youth indicator \citep[e.g.,][]{Soderblom1998}. 

The equivalent width of the lithium $\lambda$6707.8 line, $W_{\rm Li\textsc{i}}=224\pm8$\,m\AA, is the average value 
of all the EWs measured in each FOCES spectrum.	
We derived a lithium abundance $A_{\rm Li}= 3.2$ by interpolation of the NLTE curves of growth tabulated by \citet{PavMag96}.
A contribution of 9\,m\AA\ due to the Fe{\sc{i}} $\lambda$ 6707.4 \AA~line, evaluated according to the empirical correction 
 proposed by \cite{Soderblom1993}, $\Delta EW_{\rm Li} ({\rm m\AA})=20\,(B-V)_0-3$, was subtracted before calculating the abundance. 
Although the lithium content cannot be simply converted into star age, HD~171488 seems to be slightly more lithium-rich than 
the stars in Pleiades upper envelope of the same temperature. The lithium abundance indicates an age in the range 50--150 Myr, 
in agreement with the position on the HR diagram.

\subsection{Metallicity}
\label{sec:metall}

Iron abundance was measured, in the local thermodynamic equilibrium (LTE) assumption, using the 2002 version
of \textsc{Moog} \citep{sneden1973}  and a grid of 1D model atmospheres by \citet{Kuru93}. LTE iron abundance was derived by means 
of equivalent widths of 31 \ion{Fe}{i} and \ion{Fe}{ii} lines in the 5000--6800\,\AA~range measured using a Gaussian 
fitting procedure with the IRAF task \textsc{Splot}. 
For the analysis, we used the prescriptions given by \citet{Randich2006}. We refer to that paper for a wide description 
of the procedure, line list, and $\sigma$-clipping criteria.

The effective temperature was determined by imposing that the iron abundance does not depend on the excitation
potentials of the lines. The microturbulence velocity $\xi$ was determined by imposing that the iron abundance
is independent of the  equivalent widths of \ion{Fe}{i} lines. The surface gravity $\log g$ was determined by 
imposing the \ion{Fe}{i}/\ion{Fe}{ii} ionization equilibrium.
The initial value for the effective temperature was the one we adopted as undisturbed photospheric $T_{\rm eff }=5800$\,K  
(Sect.~\ref{sec:spec_type}). For the surface gravity, we adopted as starting value the one derived by means of the \textsc{Rotfit} code for 
spectral synthesis ($\log g=4.3$; Sect.~\ref{sec:spec_type}). 
The initial microturbulence velocity was set to be 1.5 km\,s$^{-1}$. Final astrophysical parameters were 
$T_{\rm eff}=5850\pm100$\,K, $\log g=4.3\pm0.2$, $\xi=1.8\pm0.2$\,km\,s$^{-1}$, i.e. the effective temperature and gravity
did not change appreciably, while the iron abundance was [Fe/H]=+0.15$\pm$0.07 which is in agreement, taking the errors into
account, with the value provided by \textsc{Rotfit}, but it is noticeably higher than the value of $-0.5$ suggested by \citet{Strass03}.

\section{Photospheric activity}
\label{sec:phot_act}

\subsection{Light curve and spot modeling}
The photometric data acquired contemporaneously to the spectroscopic ones allowed us to obtain a light-curve 
showing a rotational modulation due to the presence of spots 
(Fig.~\ref{fig:modulations}). The rotational phases were derived according to the \citet{Jarvi08} ephemeris:
\begin{equation}
HJD_{\phi=0} = 2\,449\,950.550+1.^{\rm d}33697\,\times\,E\,,
\label{Eq:ephem_lambda}
\end{equation}
{\noindent where the rotational period ($P_{\rm rot}$) is nearly identical to the value of 1.3371 days found by \cite{Strass03}. 
With a periodogram analysis and a {\sc clean} deconvolution algorithm (\citealt{Roberts}) 
we find a period of $1.37\pm0.12$ days that is fully consistent with the value of \citet{Jarvi08} taking the limited time span of our 
data into account.}

The $V$ light-curve has a slightly asymmetric shape with a maximum brightness $V_{\rm max}\simeq 7.^{\rm m}40$ at about
phase 0.4, a minimum around 0.0 phase, and a variation amplitude $\Delta V\simeq$0.07 mag. 
Its shape is very similar to that shown by \citet{Jarvi08} in their Fig.~2 for the mean epoch 2006.79.

For a first reconstruction of the photospheric inhomogeneities, we  used \textsc{Macula} \citep{Frasca05}, a spot model code 
which assumes circular dark or bright spots on the surface of a spherical limb-darkened star.
The flux contrast between spotted areas and quiet photosphere ($F_{\rm sp}/F_{\rm ph}$) is evaluated through 
the PHOENIX NextGen \citep{Haus99} atmosphere models adopting $T=5800$\,K for the quiet photosphere.
With only two photometric bands and the rather small amplitude of the light curve it is not easy to derive the 
spot temperature. Thus we preferred to fix the spot temperature, $\Delta T=1500$\,K, close to the values found for high 
latitude spots by \citet{Strass03} and \citet{Jarvi08} with DI technique, and let the spot area and
location be free to vary. In order to minimize the degrees of freedom and still allowing the code to reproduce 
an asymmetric light curve, only two spots have been put on the photosphere. We adopted the same inclination of the rotation axis 
used by \citet{Strass03}, $i=55^\circ$. We also exploited the information contained in the radial velocity searching for a solution
of the $B$, $V$ light curves that simultaneously reproduce the $V_{\rm r}$ behaviour. 
The synthetic curves are displayed by dotted lines superimposed to the data in Fig.~\ref{fig:modulations} and the spot 
parameters are quoted in Table~\ref{tab:spotplage}.
The light curves alone mainly constrains the spot longitudes and areas, with only a 
raw indication of latitudes. Anyway, the broad light modulation, as well as the shape and amplitude of radial velocity 
variations, force the spots to have high latitudes ($75^\circ$ and $60^\circ$) as found in previous studies.
This simple model will be also applied to the chromospheric line-fluxes curves in Section~\ref{sec:discussion} for the
study of the spatial relation between photospheric and chromospheric active regions (ARs).

\subsection{Doppler Imaging}
\label{sec:DI}

\begin{figure*}[th]
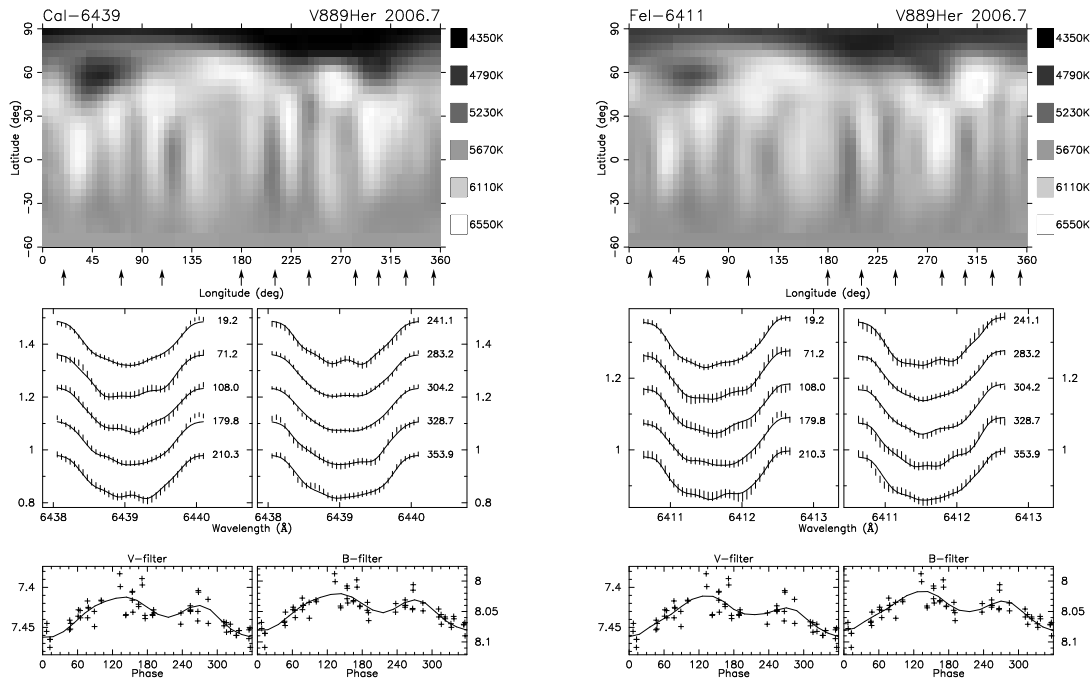

\centering
\hspace{0cm}
\includegraphics[width=6.5cm,clip]{14460f2a.ps}
\hspace{1cm}
\includegraphics[width=6.5cm,clip]{14460f2b.ps}
\caption{Doppler maps ({\it top panels}), line-profile fits ({\it middle panels}), and light-curve fits ({\it bottom panels}) for 
Ca\,{\sc i}-$\lambda$6439  ({\it left panels}) and Fe\,{\sc i}-$\lambda$6411 ({\it right panels}). The arrows below each map indicate 
the phases of spectroscopic observations.} 
\label{fig:doppler_maps}
\end{figure*}

For a more accurate surface reconstruction we used the Doppler imaging code
{\sc TempMap} by \citet{Rice89}.
It performs a full LTE spectrum synthesis by solving the equation of transfer 
through a set of model atmospheres \citep{Kuru93} at all aspect angles and for 
a given set of chemical abundances.
Simultaneous inversions of the spectral lines as well as of two photometric 
band-passes (Johnson $B$ and $V$ in the present case) are then carried out using 
maximum-entropy regularization.

The 10 available spectra collected over 3 rotations cover the whole rotational phase, i.e., they can be 
used to reconstruct one single Doppler image.
The last spectrum from the time series was excluded due to an insufficient $S/N$ ratio.

\begin{figure}[bht]
\centering
\includegraphics[width=8.5cm,clip]{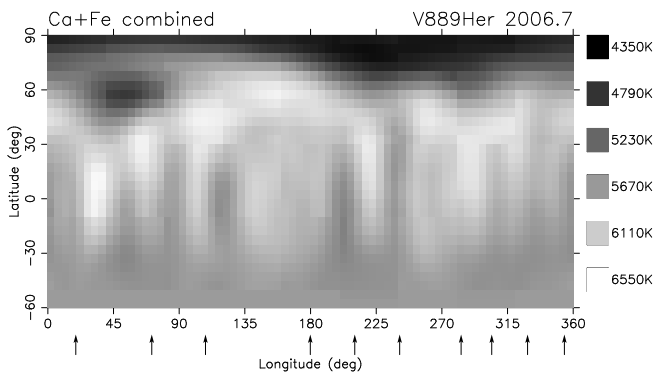}	
\caption{Average map from the two single-line inversions in Fig.~\ref{fig:doppler_maps}.} 
\label{fig:averagemap}
\end{figure}

Doppler imaging was performed within the 6392--6440~\AA\ spectral range using two lines, the Fe\,\,{\sc i}-6411, 
and Ca\,\,{\sc i}-6439 lines. The quality of the data with a low $S/N$ ratio did not allow us to use other lines in the domain. 
The individual reconstructions, as well as the average map, are shown in Fig.~\ref{fig:doppler_maps} and Fig.~\ref{fig:averagemap},
respectively. The maps revealed a similar spot distribution, i.e., 
mainly cool polar spots with temperature contrasts of up to $\approx$1500\,K with respect to the undisturbed surface of 5800\,K.
Some low-latitude features are also recovered, however, with significantly weaker contrast ranging from $\approx$300\,K (Ca\,\,{\sc i}-6439)  
to a maximum of $\approx$500\,K (Fe\,\,{\sc i}-6411). Numerous bright features are also appeared, as mirroring of dominant cool spots, thus 
most of them are believed to be artifacts.

Despite small differences, the resulting Ca and Fe maps are in very good agreement (cf. the average map in 
Fig.~\ref{fig:averagemap}). The overall structure reveals a dominant cool polar spot with a broad appendage extending in 
longitude between $\sim180^{\circ}$--$300^{\circ}$
and another cool region, at a mid-latitude of $\sim60^{\circ}$, centered at a longitude $\phi\simeq 45^{\circ}$.
This result is reminiscent of the maps obtained by \citet{Jarvi08}, where a similar temperature distribution was recovered over 
several years.
Moreover, the polar spot/appendage is consistent with the high-latitude circular spot (\#\,2) found by modelling our light curve with  
\textsc{Macula} while the \#\,1 region found by the light-curve model at $\phi=10^{\circ}$ (see
Table~\ref{tab:spotplage}) matches well the lump of the polar spot at $\sim0^{\circ}$ phase along with the mid-latitude spot at $\phi=45^{\circ}$.

\section{Chromospheric activity}
\label{sec:chrom_act}

\begin{table*}[t]   
\caption{Radial velocities and parameters of the subtracted spectra.}
\label{tab:eqw_lines}
\begin{center}
\begin{tabular}{ccccccccc}
\hline
\hline
  \noalign{\smallskip}
HJD & Phase & $V_{\rm r}$  & $W^{\rm em}_{\rm H\alpha}$ & $W^{\rm em}_{\rm H\beta}$ & $W_{\rm He\textsc{i}}$ & $W^{\rm em}_{\rm Ca\textsc{ii}-IRT}$ & 
$W^{\rm em}_{\rm Ca\textsc{ii}-H+K}$ \\ 
(+2\,400\,000) & & (km\,s$^{-1}$) & (\AA) & (\AA) & (\AA) & (\AA) & (\AA) \\ 
  \noalign{\smallskip}
\hline
  \noalign{\smallskip}
  53\,961.346 & 0.914 & $-23.16\pm$0.58 &  0.528$\pm$0.061  &  0.136$\pm$0.026  &  $0.044\pm$0.020  & 1.109$\pm$0.073 &  0.733$\pm$0.076   \\
  53\,961.439 & 0.984 & $-23.37\pm$0.57 &  0.575$\pm$0.066  &  0.141$\pm$0.024  &  $0.054\pm$0.021  & 1.140$\pm$0.057 &  0.764$\pm$0.077   \\
  53\,961.533 & 0.055 & $-23.49\pm$0.55 &  0.562$\pm$0.058  &  0.161$\pm$0.022  &  $0.067\pm$0.028  & 1.165$\pm$0.064 &  0.779$\pm$0.111   \\
  53\,962.355 & 0.670 & $-23.33\pm$0.57 &  0.536$\pm$0.066  &  0.142$\pm$0.022  &  $0.060\pm$0.023  & 1.144$\pm$0.051 &  0.776$\pm$0.081   \\
  53\,962.513 & 0.787 & $-23.49\pm$0.49 &  0.469$\pm$0.036  &  0.118$\pm$0.015  &  $0.062\pm$0.012  & 1.065$\pm$0.040 &  0.722$\pm$0.054   \\
  53\,962.589 & 0.845 & $-23.21\pm$0.53 &  0.480$\pm$0.041  &  0.152$\pm$0.032  &  $0.064\pm$0.020  & 1.076$\pm$0.053 &    ...  \\
  53\,963.465 & 0.500 & $-22.53\pm$0.57 &  0.488$\pm$0.042  &  0.131$\pm$0.018  &  $0.059\pm$0.026  & 1.009$\pm$0.058 &  0.697$\pm$0.054   \\
  53\,963.577 & 0.583 & $-23.00\pm$0.60 &  0.484$\pm$0.086  &  0.164$\pm$0.027  &  $0.085\pm$0.032  & 1.032$\pm$0.095 &  0.709$\pm$0.109   \\
  53\,964.397 & 0.197 & $-23.67\pm$0.56 &  0.593$\pm$0.082  &  0.144$\pm$0.024  &  $0.052\pm$0.027  & 1.123$\pm$0.077 &  0.765$\pm$0.070   \\
  53\,964.536 & 0.301 & $-23.46\pm$0.53 &  0.550$\pm$0.061  &  0.112$\pm$0.030  &  $0.078\pm$0.023  & 1.095$\pm$0.085 &  0.706$\pm$0.084   \\
  53\,964.612 & 0.358 & $-23.17\pm$0.61 &  0.511$\pm$0.109  &  0.149$\pm$0.039  &  $0.091\pm$0.053  & 1.114$\pm$0.143 &    ...  \\
  \noalign{\smallskip}
\hline
\end{tabular}
\end{center}
\end{table*}
\normalsize

The wide wavelength range of FOCES allowed us to study the chromosphere of HD~171488 by using several 
lines from the near UV to the NIR wavelengths (namely, Ca\,{\sc ii} H\,\&\,K, H$\beta$, He\,{\sc i}\,D$_3$, H$\alpha$, Ca\,{\sc ii} IRT), 
which provide information on different atmospheric levels, from the region of temperature minimum to the upper chromosphere. 
To derive the chromospheric losses, we used the ``spectral synthesis" technique, based on the comparison between the target 
spectrum and an observed spectrum of a non-active standard star (called ``reference spectrum"). The difference between the 
observed and the reference spectrum provides, as residual, the net chromospheric line emission, 
which can be integrated to produce an emission equivalent width, $W^{\rm em}$ \citep[see, e.g.,][]{Herb85, Bar85, Fra94, Montes95}. 

\begin{figure*}
\centering
\includegraphics[width=16cm,height=10cm]{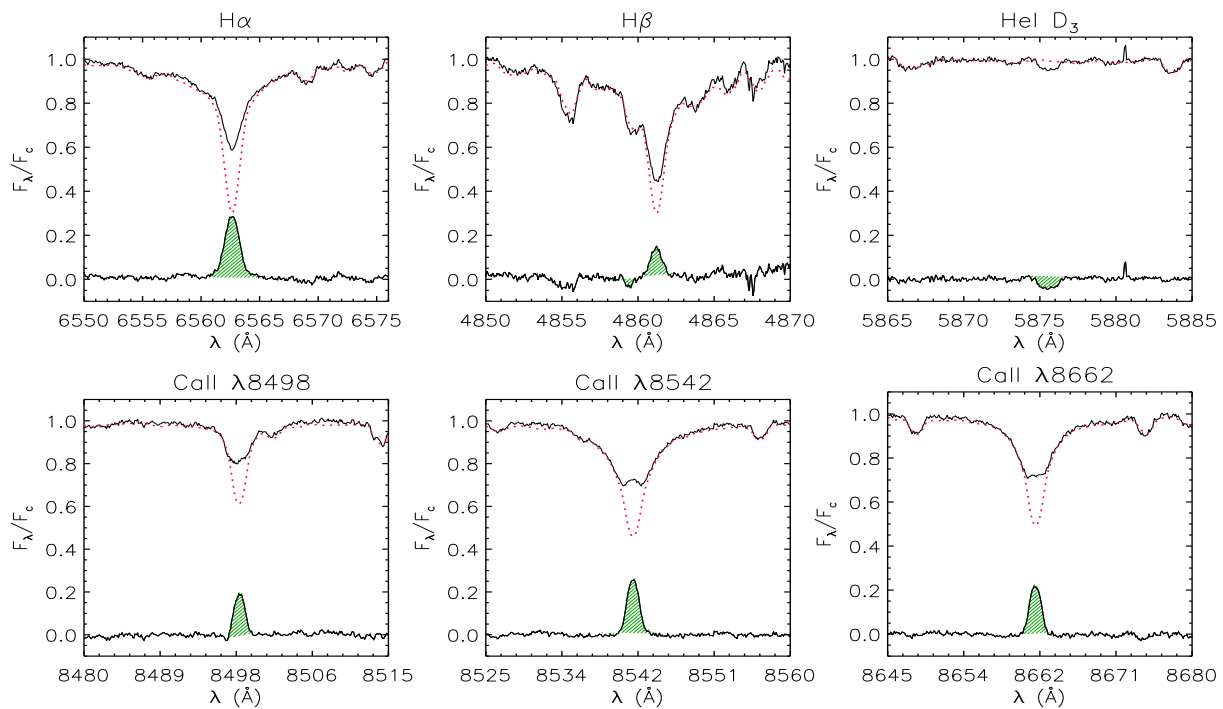}	
\caption{{\it Top of each panel}: Examples of observed, continuum-normalized spectra of HD~171488 
(solid line) in the H$\alpha$, H$\beta$, He\,{\sc i}\,D$_3$, and Ca\,{\sc ii} IRT regions together with the non-active 
stellar template (dotted line). {\it Bottom of each panel}: The difference spectra of the two upper spectra.
The hatched area represents the excess emission (absorption for He\,{\sc i}\,D$_3$) that has been integrated to 
get the net equivalent width.} 
\label{fig:Halpha_spectra}
\end{figure*}

\begin{figure*}
\centering
\includegraphics[width=13cm,height=11cm]{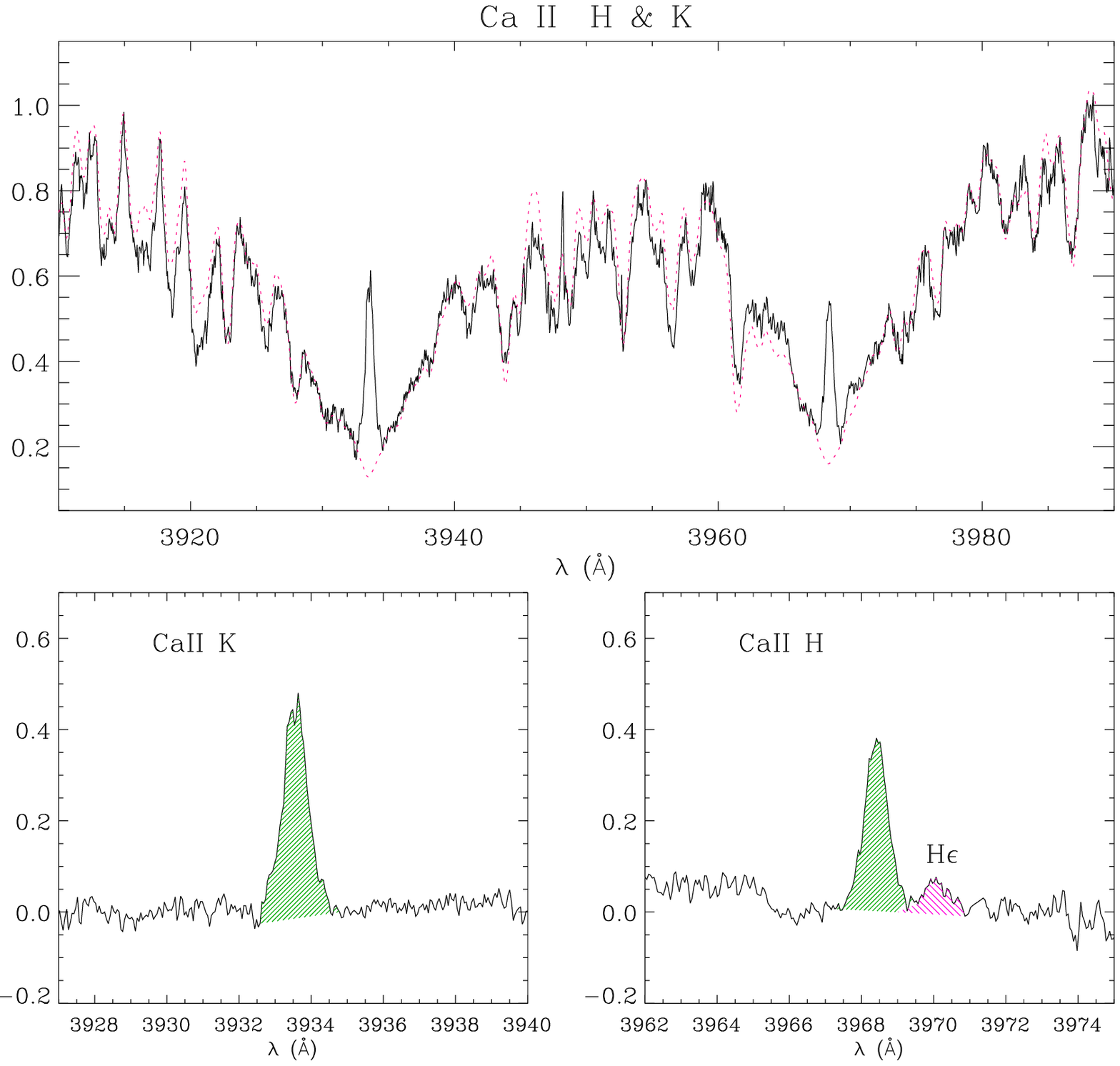}	
\caption{{\it Top panel}: Example of an observed, continuum-normalized spectrum of HD~171488 (solid line) 
in the Ca\,{\sc ii} H\,\&\,K region together with the non-active stellar template (dotted line). 
{\it Bottom panels}: The difference spectrum. H$\epsilon$ emission is barely
visible in the raw spectrum, but it is clearly emphasized in the residuals.} 
\label{fig:CaIIHK}
\end{figure*}

The non-active star used as a reference for the spectral subtraction is \object{72~Her} (=\object{HD~157214}), a G2\,V star 
($B-V=0.62$) with a very low activity level, as indicated by the low value of the Ca\,{\sc ii} Mt. Wilson index 
($S=0.156$, \citealt{Duncan91}), close to the value for the quiet Sun ($S=0.159$, \citealt{Oranje83}), and by the deepest H$\alpha$ 
core among the G-type stars investigated by \citet{Herb85}. 
This star was also observed with FOCES during the same run as HD~171488. 
In Fig.~\ref{fig:Halpha_spectra}, we show an example of a spectrum of HD~171488 in the H$\alpha$, H$\beta$, He\,{\sc i}\,D$_3$, 
and Ca\,{\sc ii} IRT regions, 
together with the standard-star spectrum rotationally broadened to $v\sin i=37$ km\,s$^{-1}$ 
which mimics the active star in absence of chromospheric activity. The Ca\,{\sc ii} H\,\&\,K region is displayed in 
Fig.~\ref{fig:CaIIHK}. The H$\alpha$, H$\beta$, and Ca\,{\sc ii} IRT profiles are clearly filled-in by emission.
The Ca\,{\sc ii} IRT $\lambda$8542 line displays a small emission reversal in its core.
The He\,{\sc i} $\lambda$5876 line appears as an absorption feature.
The Ca\,{\sc ii} H\,\&\,K cores exhibit strong emission features and H$\epsilon$ 
emission is also visible (Fig.~\ref{fig:CaIIHK}). 

The values of emission equivalent widths, $W^{\rm em}$, in the different chromospheric diagnostics are listed in 
Table~\ref{tab:eqw_lines} along with their errors.

We also evaluated the total radiative losses in the chromospheric lines following the guidelines by \citet{Fra94}, i.e. 
by multiplying the average $W^{\rm em}$ by the continuum surface flux at the wavelength of the line.	
The latter was evaluated for all the stars of the spectrophotometric atlas 
of \citet{GunnStryker} with the angular diameters calculated through the \citet{Barnes1976} relation. The continuum 
flux of HD~171488 was found by interpolating the values for the \citet{GunnStryker} stars at $B-V=0.62$. 
As a check, we also used the synthetic low-resolution NextGen spectra \citep{Haus99} 
by interpolating the tabulated continuum fluxes at the star temperature. We found nearly identical values
of the continuum flux (within 5\%) with the two methods.
The average chromospheric line fluxes are reported in Table~\ref{tab:fluxes}.

\begin{table}
\caption{Radiative chromospheric losses, $\chi^2$, and probability for the line-flux modulations.}
\centering
 \begin{tabular}{lcrrrr}
  \hline\hline
  \noalign{\smallskip}
  Line                       & Flux                       & $\chi^2_0$  & $P_{\chi^2<\chi^2_0}$ & $\chi^2_0$  & $P_{\chi^2<\chi^2_0}$ \\  			     
                & \scriptsize{(erg\,cm$^{-2}$\,s$^{-1}$)} & \multicolumn{2}{c}{Fourier pol.}  & \multicolumn{2}{c}{Constant}  \\
  \noalign{\smallskip}
  \hline
  \noalign{\smallskip}
  H$\alpha$               &  3.8$\times10^6$  & 0.32 & 0.3\,\% & 6.38 & 30\,\% \\
  H$\beta$                &  1.4$\times10^6$  & 2.22 & 10\,\%  & 5.20 & 12\,\% \\
  \ion{He}{i} D$_3$       &  ...	      & 1.53 &  4\,\%  & 2.77 &  4\,\% \\
  \ion{Ca}{ii} H+K        &  6.5$\times10^6$  & 0.02 & 0.05\,\% & 1.21 & 1\,\%  \\
  \ion{Ca}{ii} IRT        &  5.3$\times10^6$  & 0.27 & 0.2\,\% & 5.18 & 18\,\% \\
  \noalign{\smallskip}
  \hline
\end{tabular}
\label{tab:fluxes}
\end{table}

For evaluating the significance of the $W^{\rm em}$ modulations we used simple $\chi^2$ tests.
If the data are totally uncorrelated with the rotational phase, no significant $\chi^2$ difference is
expected between a fit of a smooth periodic function and that of a constant function (the weighted average of the data).
As a fitting function, we used a Fourier polynomial of degree 2 which is able 
to reproduce asymmetrical curves. Additionally, we calculated the probability $P_{\chi^2<\chi^2_0}$ that, in a $\chi^2$ distribution 
with $n-p-1$ degrees of freedom (with $n$ the number of data points and $p$ the number of fitting parameters), a random variable is 
less than the measured $\chi^2_0$.  A small value of $P_{\chi^2<\chi^2_0}$ for the periodic function assures a statistically significant 
modulation. The $\chi^2$ and $P_{\chi^2<\chi^2_0}$ values are also reported in Table~\ref{tab:fluxes}.

\subsection{H$\alpha$ line}
The H$\alpha$ line is one of the most useful and easily accessible indicators of chromospheric emission related to solar and 
stellar activity. 
Its source function is photoionization-dominated in the quiet chromospheres of the Sun and 
solar-like stars, but it can become collision-dominated in very active stars or inside ARs.	
As a consequence, H$\alpha$ is useful for detecting chromospheric solar and stellar plages, thanks to their 
high contrast against the surrounding chromosphere.

All our FOCES spectra of HD~171488 always show an H$\alpha$ absorption profile with a considerable filling-in of the core. 
The residual H$\alpha$ profile is almost symmetric (see Fig.~\ref{fig:Halpha_spectra}). 
The values of the residual emission $W^{\rm em}_{\rm H\alpha}$ 
are plotted as a function of the rotational phase in Fig.~\ref{fig:modulations}, where
a smooth rotational modulation anti-correlated with the light curve is apparent.
The results of the $\chi^2$ analysis, summarized in Table~\ref{tab:fluxes}, assure the statistical significance of the observed 
H$\alpha$ rotational modulation.

\subsection{H$\beta$ line}

When compared with the non-active template, the H$\beta$ profile of HD~171488 also appears  filled in by core emission. 
This contribution is, however, much smaller than for the H$\alpha$ line, as expected from the lower transition probabilities 
for H$\beta$ and the higher photospheric flux at 4861\,\AA.

The results of the $\chi^2$ analysis (Table~\ref{tab:fluxes}) indicate that the small-amplitude variation displayed 
by $W^{\rm em}_{\rm H\beta}$ is not statistically significant.

As a further test, we calculated the Spearman's rank-correlation coefficient between $W^{\rm em}_{\rm H\beta}$ and $W^{\rm em}_{\rm H\alpha}$ 
by means of the IDL procedure {\sc r\underline{ }correlate} \citep{Press86}. We found a very low correlation coefficient 
$\rho=6\times10^{-8}$ with a significance $1.0$, which means no correlation.
We remark that this could be due to the faintness of the residual emission that prevents us from clearly detecting variations over the errors.

On the basis of the H$\alpha$ and H$\beta$ flux we evaluated a Balmer decrement
$F_{\rm H\alpha}/F_{\rm H\beta}\,\simeq\,2.7\,\pm\,0.6$.
As suggested by \citet{Hall92}, high values of the Balmer decrement ($\sim 10$) are typical of solar prominences 
seen off-limb (e.g., \citealt{LandMong79}), whereas solar plages or flares have always values of about 1--2 (e.g., \citealt{Chester91}). 
Indeed, the theoretical NLTE model developed by \citet{Buza89} leads to high Balmer decrement values, ranging from
3 to 15 for prominence-like structures viewed away from the stellar disk, while values in the range 1--2 are found for plage-like
structures. 

 The Balmer decrement measured for HD~171488 suggests that the emission in the Balmer lines is basically due to magnetic surface 
regions analogue to solar plages and eventual prominences play a marginal role.

\subsection{Helium D$_{3}$ line}

The He\,{\sc i} $\lambda$\,5876 (D$_{3}$) line is another useful diagnostics of chromospheric activity in the optical
domain. It is particularly helpful for F-type stars in which the other activity indicators in the optical/IR range 
are more difficult to be observed due to the strong continuum flux \citep[e.g.,][]{Rachford09}. Both in these stars and in the 
moderately active G an K-type stars, this line is normally seen as an absorption feature \citep[e.g., ][]{Huene86,Biazzo07b}.
The He\,{\sc i} absorption implies a temperature of $\sim$ 10\,000\,K in the layer where the line is formed, i.e. this line,
as in the Sun, is a diagnostics of the upper chromosphere. 

The adopted reference spectrum (72~Her) does not show any He\,{\sc i} absorption nor emission, as expected on the basis of
its very low activity level.
In this case, the spectral subtraction technique, cleaning the spectrum from nearby photospheric absorption lines, 
allows to emphasize the He\,{\sc i} line of HD~171488, which appears as an excess absorption in the residual spectra (see 
Fig.~\ref{fig:Halpha_spectra}), and to measure its equivalent width, $W_{\rm HeI}$.
For $W_{\rm HeI}$ we adopted the usual convention that an absorption line has a positive equivalent width.

As seen in Table~\ref{tab:eqw_lines} and Fig.~\ref{fig:modulations}, the D$_{3}$ line is always in absorption with 
small variations only possibly correlated with the star rotation.

As for the H$\beta$, the $\chi^2$ test indicates that there is no significant rotational modulation of $W_{\rm HeI}$.
However, the rank-correlation coefficient with the H$\alpha$, $\rho=-0.345$ with a significance $0.298$, indicates a
marginal anti-correlation between $W^{\rm em}_{\rm H\alpha}$  and $W_{\rm HeI}$, in the sense that the  He\,{\sc i} absorption 
is slightly weaker when the net H$\alpha$ emission gets stronger.

\subsection{Ca\,{\sc ii} H\&K lines}

The {Ca}{\sc ii} H\,\&\,K lines show strong emission cores that appear nearly symmetric in all spectra without any 
detectable trace of self-absorption. H$\epsilon$ emission is barely seen only in the few best 
spectra and become more evident after the subtraction of the non-active template (Fig.~\ref{fig:CaIIHK}).  
Thus, we could not investigate the behaviour of H$\epsilon$ with the star rotation as we did for SAO~51891 \citep{Biazzo09}.

The absorption wings of each of the two Ca\,{\sc ii} lines span two {\it \'echelle} orders and it was necessary 
to merge them, following the guidelines of  \citet{Frasca00}, before proceeding with the spectral subtraction analysis. 
The non-active template was built up, with the same merging procedure, from the FOCES spectrum of 72~Her broadened at
$v\sin i=37.1$\,km\,s$^{-1}$. No emission in the core of the Ca\,{\sc ii} H and K lines of the template is visible 
(Fig.~\ref{fig:CaIIHK}, upper panel), as expected from the very low chromospheric activity level reported in the literature (see above).

We measured the equivalent widths by integrating the emission profiles in the subtracted spectra, as for H$\alpha$.
We disentangled the contribution of the  H$\epsilon$ line from the {Ca}\,{\sc ii} H by fitting Voigt functions to
both profiles.
The sum of the net equivalent widths of H and K lines, $W^{\rm em}_{{\rm Ca}\textsc{ii}-H+K}$, displays a significant 
modulation with the star rotation (bottom left panel in Fig~\ref{fig:modulations}) that is not as outstanding as that
of H$\alpha$ but its reliability is testified by the results of the $\chi^2$ tests (Table~\ref{tab:fluxes}).
The rank-correlation coefficient, $\rho=0.567$ with a significance $0.112$, supports a rotational modulation of the
Ca\,{\sc ii} H+K emission correlated with the H$\alpha$ one.

We calculated the radiative chromospheric losses in the H\,\&\,K lines analogously as done for H$\alpha$.
In particular, we used two 10\,\AA-wide bands centered at 3910 and 4010~\AA, i.e. at the two sides of the H\,\&\,K 
lines, to perform the flux calibration, following the prescriptions by \citet{Frasca00}. 
The ratio of the residual peak intensities,  $r_{\rm K}/r_{\rm H}\simeq\,1.08\pm0.015$, is larger than typical
values measured in the quiet solar chromosphere ($r_{\rm K}/r_{\rm H}\simeq$1.05) but it is smaller than what it is 
found in solar plages ($r_{\rm K}/r_{\rm H}\simeq$1.20, e.g. \citealt{Linsky70}). If the solar analogy is valid, this 
would mean that we are observing, at all phases, a ``mixture'' of quiet chromosphere and ARs.	

\subsection{Ca\,{\sc ii} IRT lines}

The lines of the {Ca}{\sc ii} infrared triplet, which share the upper level of the H and K transitions, present some 
advantages compared to the {Ca}{\sc ii} H\,\&\,K lines. They lie in a spectral region with a well-defined continuum, 
making the normalization easier during the data reduction. Moreover, they are not significantly affected by telluric 
lines and are less affected by atmospheric extinction than the visible and ultraviolet lines.
As such, they have become very useful chromospheric diagnostics in recent years also thanks to the high sensitivity of
the new detectors to the near infrared. 

A filled-in {Ca}{\sc ii} IRT-3 ($\lambda 8862$) line profile is shown by \citet{Busa07} which measured a net equivalent width 
(after the subtraction of a synthetic profile) of 0.511 \AA\  that is larger than the values measured by us, ranging from 
0.350 to 0.402 \AA. The total contribution of the three IRT lines measured by \citet{Busa07} in 2002 is 1.46 \AA, which 
is significantly larger than the values measured by us (1.01--1.16\,\AA, see Table~\ref{tab:eqw_lines}). This could be due to a 
higher activity level during the Bus\`a et al. observations or, more likely, to the different photospheric template used by them, 
that is apparently deeper in its core than our own.

Measurements of the residual EW of {Ca}\,{\sc ii} IRT-3 lines (after the subtraction of a solar spectrum broadened at the 
star's $v\sin i$) are also reported by \citet{Jarvi08} for 2001, 2002, and 2005 epochs.
They found values comparable to our determinations in 2001 and 2002 and a larger EW, on average, in 2005.
Moreover, they detected also a modulation of the residual EW anti-correlated with the contemporaneous light 
curve in the last season.

\begin{figure*}
\centering
\includegraphics[width=6.cm,height=8cm]{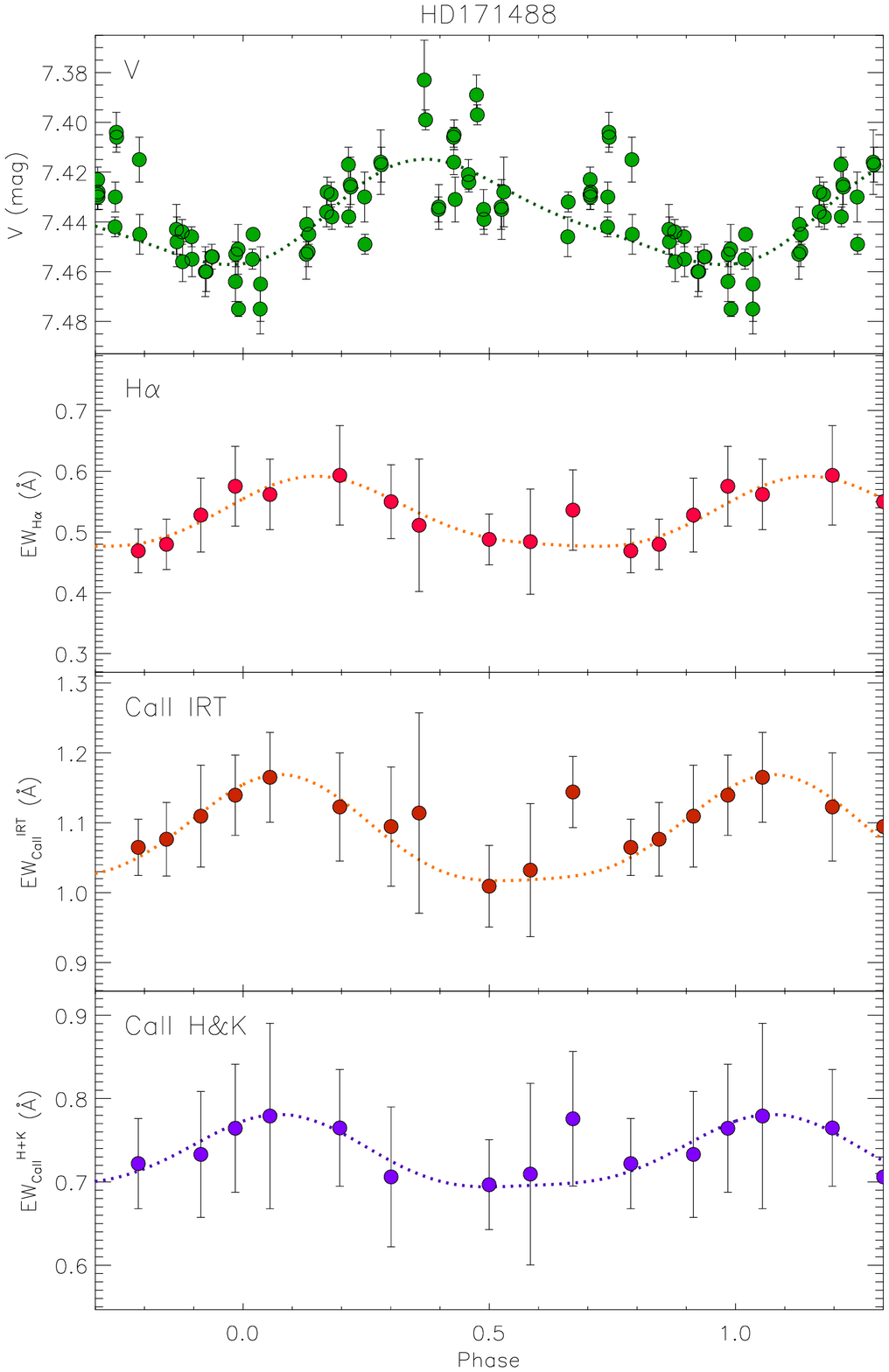}	
\hspace{1cm}
\includegraphics[width=6.cm,height=8cm]{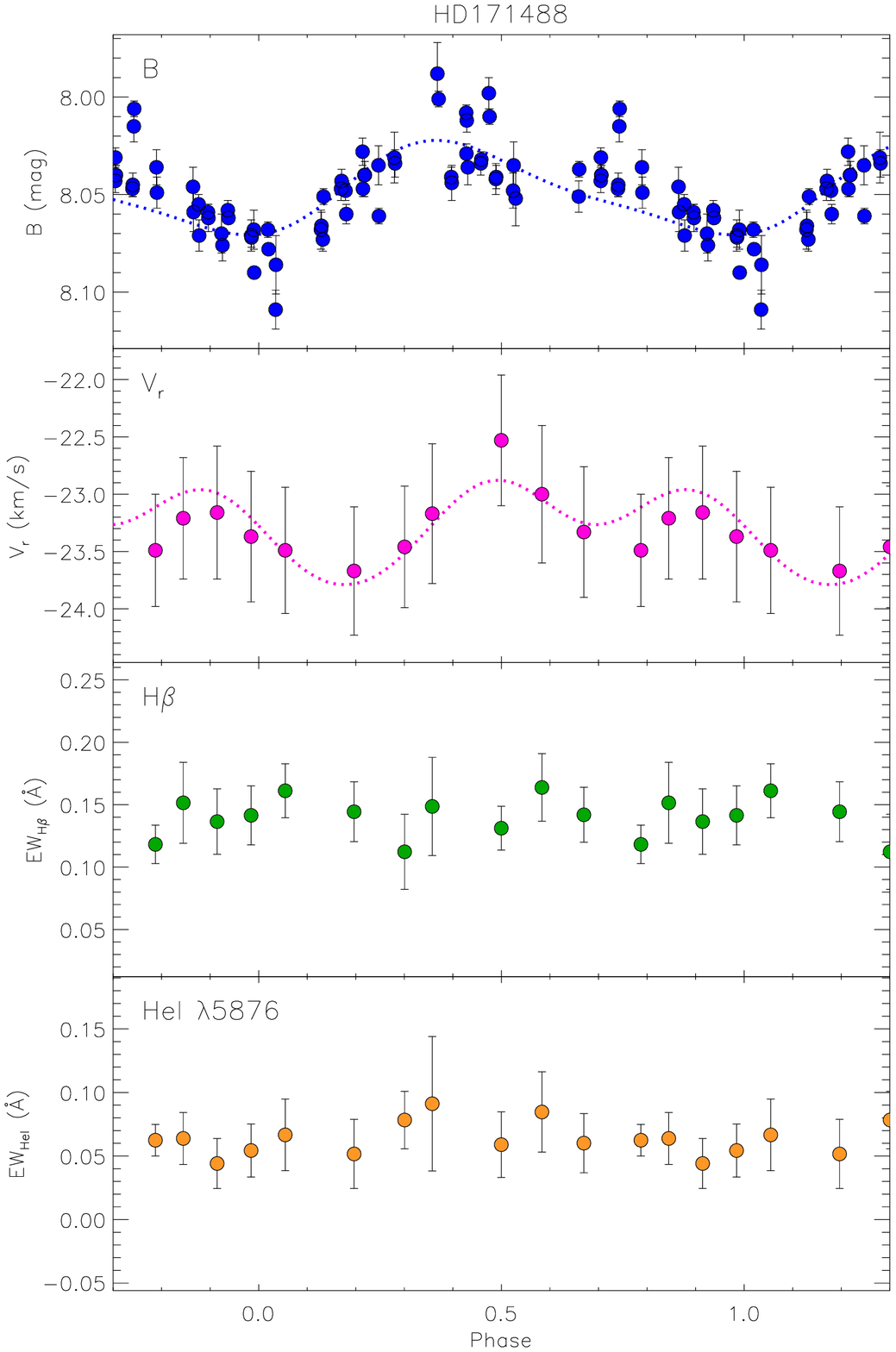}	
\caption{{\it Left panels. From top to bottom.}  $V$ magnitude, $W^{\rm em}_{\rm H\alpha}$,  $W^{\rm em}_{\rm Ca\textsc{ii}-IRT}$,
and $W^{\rm em}_{\rm Ca\textsc{ii}-H+K}$ versus the rotational phase. Note the clear anti-correlation of the photospheric 
($V$ light curve) and chromospheric diagnostics. The dotted lines represent the solution of spot/plage model. 
{\it Right panels. From top to bottom.} $B$  magnitude, radial velocity,
$W^{\rm em}_{\rm H\beta}$, and $W_{\rm He\textsc{i}}$ versus the rotational phase.  } 
\label{fig:modulations}
\end{figure*}

\begin{figure}
\centering
\includegraphics[width=9cm]{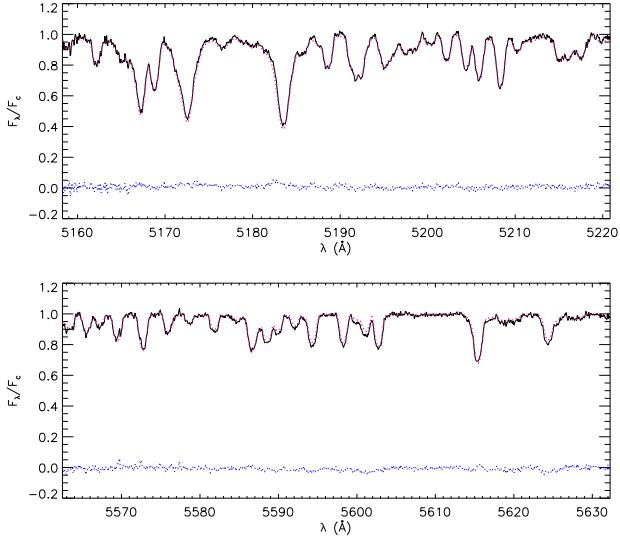}	
\caption{Observed FOCES spectrum of HD~171488 in the Mg\,{\sc i}b triplet (upper panel) and 5600 {\AA} (lower panel) 
spectral regions together with the ELODIE standard spectrum broadened at 37.1 km s$^{-1}$ overplotted with a thin/red line. 
In each box the difference (observed $-$ synthetic) is also displayed with a dotted line around zero ordinate. } 
\label{fig:spe}
\end{figure}

We found a nice rotational modulation of the net equivalent widths of the IRT lines, which becomes more evident if we consider 
the total emission of the triplet $W^{\rm em}_{\rm Ca\textsc{ii}-IRT}$ (see Fig.~\ref{fig:modulations}) that closely follows the
H$\alpha$ trend. 
The statistical significance of the rotational modulation is assured both from the $\chi^2$ analysis (Table~\ref{tab:fluxes})
and from the strong correlation with the $W^{\rm em}_{\rm H\alpha}$ curve ($\rho=0.755$ with a significance $0.007$).

The flux ratio $F_{8542}/F_{8498}=1.7$, indicative of large optical depths, is in the range of values (1.5--3) found by 
\citet{Chester91} in solar plages. Solar prominences have instead values of $F_{8542}/F_{8498}\sim 9$, typical of an 
optically-thin emission source.  This result confirms what we found for HD~171488 from the H$\alpha$/H$\beta$ flux ratio.

\section{Plage modeling}
\label{sec:discussion}

The chromospheric line-flux variations contain valuable information on the surface inhomogeneities at this atmospheric level
simultaneous to the photospheric diagnostics.

We have applied our spot-model IDL code, \textsc{Macula} \citep{Frasca05}, to the curves of $W^{\rm em}_{\rm H\alpha}$, 
$W^{\rm em}_{\rm Ca\textsc{ii}-IRT}$, and $W^{\rm em}_{\rm Ca\textsc{ii}-H+K}$, with the aim of further investigating the 
main parameters of the chromospheric plages and the degree of spot-plage correlation. 
Given the scatter in the data, two bright spots (plages) are  fully sufficient to reproduce the observed variations. 

An important parameter that enters in the plage model is the flux contrast between plages and surrounding chromosphere 
($F_{\rm plage}/F_{\rm chrom}$). Values of $F_{\rm plage}/F_{\rm chrom}$ $\approx\,2$, that can be deduced averaging 
solar plages in H$\alpha$ \citep[e.g., ][]{Elli52,LaBo86b}, are too low for reproducing the $\sim$23\% variation of 
$W^{\rm em}_{\rm H\alpha}$. 
In fact, very large plages, covering a large fraction of the stellar surface, would be required with such a low flux 
contrast and they do not provide a good fit of the observed modulation on the basis of the $\chi^2$. 
On the other hand, very high values of flux contrast ($F_{\rm plage}/F_{\rm chrom}\,\gtsim 10$) would imply very small plages 
producing top-flattened modulations that are not observed.
In order to achieve a good fit, a flux contrast of 5 was adopted.

The solutions essentially provide the longitude of the plages, giving only rough estimates of their latitude and size. 
We searched for the best solution starting with the values of longitude, latitude, and  radii of the photospheric active 
regions and we found that only the longitude of the plage at lower latitude must change appreciably. The plage radii 
are, however, strongly dependent on the assumed flux contrast $F_{\rm plage}/F_{\rm chrom}$. This means that only the combined 
information between plage sizes and flux contrast, i.e. some kind of plage ``luminosity'' in units of the quiet 
chromosphere ($L_{\rm plage}/L_{\rm quiet}$) can be deduced as a meaningful parameter. 
We remark that the $W^{\rm em}$ level of the quiet chromosphere is also a rather uncertain parameter, because it depends on 
the plage distribution which can give a non-negligible contribution also at the phase of minimum $W^{\rm em}$. 
The solution of the H$\alpha$ light curve is displayed with a dotted line in Fig.~\ref{fig:modulations} along
with the observed $W^{\rm em}_{\rm H\alpha}$ values.

For the solution of the $W^{\rm em}_{\rm Ca\textsc{ii}-IRT}$ and $W^{\rm em}_{\rm Ca\textsc{ii}-H+K}$ curves we had to use
a lower plage contrast with the same radii used for the H$\alpha$ curve, namely $F_{\rm plage}/F_{\rm chrom}=$3.3 for
the former and and 3.1 for the latter. This is consistent with the variation amplitude of about 14\% and 11\% for 
the $W^{\rm em}_{\rm Ca\textsc{ii}-IRT}$ and $W^{\rm em}_{\rm Ca\textsc{ii}-H+K}$, respectively.
The solutions of the $W^{\rm em}_{\rm Ca\textsc{ii}-IRT}$ and $W^{\rm em}_{\rm Ca\textsc{ii}-H+K}$ light curves are also 
displayed with dotted lines in Fig.~\ref{fig:modulations} and the parameters of the spots and plages, as recovered by our model, 
are listed in Table~\ref{tab:spotplage}. 
The values of plage contrast in H$\alpha$ and Ca\,{\sc ii} are nearly the same found by \citet{Frasca00} for HD~206860,
another young ($age\sim300$ Myr) early-G type star. 

\begin{table}[htb]
\caption{Spots and plages parameters.} \label{tab:spotplage}
\begin{tabular}{llcrccc}

\hline
\hline
\noalign{\smallskip}
 Diags.  & AR & Radius  &  Lon.$^{a}$  &   Lat.  &  $\frac{\rm F_{plage}}{\rm F_{chr}}$ &
 $\frac{\rm T_{spot}}{\rm T_{phot}}$ \\ 
 \noalign{\smallskip}
\hline
\noalign{\smallskip}
 $V$  & \# 1  & 15$^{\circ}.$5   &  10$^{\circ}$   &  60$^{\circ}$ &    &  0.74 \\
      & \# 2  & 16$^{\circ}.$3   & 215$^{\circ}$   &  75$^{\circ}$ &    &  0.74 \\
 H$\alpha$  &  \# 1 & 19$^{\circ}.$2   &  50$^{\circ}$   &  60$^{\circ}$	& 5.0  &  \\ 
            &  \# 2 & 14$^{\circ}.$0  &  217$^{\circ}$   &  75$^{\circ}$	& 5.0  &  \\
 Ca~{\sc ii}-IRT & \# 1  & 19$^{\circ}.$2   &  29$^{\circ}$  &  60$^{\circ}$  & 3.3  &  \\
                 & \# 2  & 14$^{\circ}.$0   & 217$^{\circ}$  &  75$^{\circ}$  & 3.3  &  \\
 Ca~{\sc ii}-HK  & \# 1  & 19$^{\circ}.$2   &  29$^{\circ}$  &  60$^{\circ}$  & 3.1  &  \\
                 & \# 2  & 15$^{\circ}.$5   & 217$^{\circ}$  &  75$^{\circ}$  & 3.1  &  \\
\noalign{\smallskip}
\hline \\
\end{tabular}
~\\
{\small $^{a}$ Longitude increases with phase, and 0$^{\circ}$ longitude
corresponds to 0.0 phase.}	
\end{table}

Considering the aforementioned size/contrast degeneracy, this would 
imply that, in these very active solar-type stars, the plages are brighter (compared to the adjacent quiet chromosphere) 
when observed in the H$\alpha$ rather than in the Ca\,{\sc ii} K line. 
This result is not in contradiction to what is known from observations of solar ARs 
in UV emission lines, for 
which the enhancement of intensity inside plages increases with the temperature of line formation (e.g., \citealt{Noyes70}).
%
As mentioned above, the flux contrast for H$\alpha$ solar plages, as measured from  residual line-core intensity,
$F_{\rm plage}/F_{\rm chrom}\approx$1.5--2.5 \citep[e.g.,][]{Elli52,LaBo86b},  is much smaller than the value derived by 
us for HD~171488 and HD~206860, while  $F_{\rm plage}/F_{\rm chrom}\simeq3$ found for the \ion{Ca}{ii} plages falls
in the range found for the solar plages \citep[e.g.,][]{LaBo86a,Ayres86}. 
Anyway, for a proper comparison with  solar plages, we must consider that for the latter the residual photospheric 
flux in the line core is not removed, and indeed it is not easy to be distinguished from the true chromospheric emission
at the low activity level of the quiet Sun.  This photospheric contribution is higher in the core of H$\alpha$ than in the H\&K 
\ion{Ca}{ii} lines giving rise to a contrast drop.
To this aim, we made our values of the H$\alpha$ and \ion{Ca}{ii} emission equivalent widths comparable with 
the solar determinations by adding back the photospheric contribution at the bottom of the lines as measured in the non-active template.

We found a strongly reduced amplitude, $\sim12$\%, for the H$\alpha$ curve for which our model derives a plage flux contrast
of 1.8, in line with the average solar values. For the \ion{Ca}{ii} H\&K residual intensity, the modulation drops from $\sim 11$\%
to $\sim 7$\% which lowers the plage contrast from 3.1 to 2.5. This would imply that, for very active solar-type stars, 
the plage contrast (with respect to their ``quiet'' chromosphere) is analogous to what found for the Sun but the overall
chromospheric flux is much larger and the plages are bigger. This gives rise to the wide observed line-flux modulations. 

Another interesting result of our plage model is that the longitude of the plage at lower latitude (those one
giving the highest contribution in the flux curves) is larger than the one of the corresponding spot by 
about 20--40$^{\circ}$, i.e. the plage is preceding the spot in the sense of star's rotation.
This result is found in all the chromospheric diagnostics for which a significant modulation has been detected
and is confirmed by the simple cross-correlation of the light curve with the $W^{\rm em}$ modulations.
We found a deep minimum of the CCF for the correlation between $V$-light curve and $W^{\rm em}_{\rm H\alpha}$ 
with a phase shift of the H$\alpha$ with respect to the $V$ curve of 0.094 in phase units, which translates into 
a longitude difference of about +34$^{\circ}$.
Slightly smaller phase shifts are found for Ca\,{\sc ii} IRT (0.070) and Ca\,{\sc ii} H+K (0.052) corresponding to
a longitude difference of about 20$^{\circ}$ between plage and spot.

Lead and lag between plages and spots have been suggested to explain phase shifts 
between photometric light curves and chromospheric indicators that are not exactly in anti-phase. \citet{Radick87} found 
spots leading plages by about 19$^{\circ}$ in the Hyades star \object{VB~31} (G0\,V), while \cite{Cata00} found both lead 
and lag in the active components of a few RS~CVn systems, with a tendency for spots to lead plages. The Sun itself, when 
observed as a star in the integrated light of the C~{\sc ii} chromospheric line ($\lambda\,$133.5~nm) from UARS SOLSTICE 
experiment, displays alternating positive and negative longitude shifts with respect to the spot photocenters \citep{Cata98}.

\section{Conclusion}
\label{sec:concl}

We performed an accurate analysis of high-resolution spectra and contemporaneous $BV$ 
photometry of the nearby young solar-type star HD~171488 with the main aim to characterize 
the behaviour of spots and plages on its surface. Revised astrophysical parameters, such as $T_{\rm eff}$, 
$\log g$, [Fe/H], rotational and heliocentric radial velocity, and lithium abundance, 
were also determined and have been found in close agreement with previous determinations from the literature, 
with the exception of the metallicity for which we did not find a sub-solar abundance as 
suggested by \citet{Strass03}. 
Both the position on the HR diagram and the lithium abundance give an estimate of about 50\,Myr for the age of 
HD~171488, confirming it as a young Sun on its way to the ZAMS.

We were able to find a clear radial velocity modulation  with an amplitude of about 
500\,m\,s$^{-1}$ due to a ``Rossiter-McLaughlin'' effect caused by starspots.
A spot modeling of the $B$ and $V$ light curves, performed with our code \textsc{Macula},
is also able to reproduce the radial velocity modulation by adopting only two high-latitude
spots 1500\,K cooler than the quiet photosphere.  
A more accurate surface reconstruction was also performed by applying the Doppler Imaging analysis to 
two photospheric spectral lines and the $B$ and $V$ light curves. The method provides maps that 
essentially confirm the large polar spotted area with a temperature difference $\Delta T\simeq$\,1500\,K
with respect to the unspotted photosphere.

Chromospheric activity was also studied with different spectral diagnostics. Equivalent 
widths of the residual emissions in the cores of H$\alpha$, \ion{Ca}{ii} H\&K, and \ion{Ca}{ii} IRT lines 
show clear modulations with the rotational phase and are ascribed to the presence of active chromospheric regions analogue 
to solar plages.
A small but significant emission in the H$\beta$ core, as well as an absorption feature corresponding to the \ion{He}{i} 
$\lambda5876$ line, are emphasized by the subtraction of the non-active template and are additional proofs of the 
high activity level of HD~171488 also in the upper chromosphere. Due to the fairly large relative errors, no clear 
rotational modulation could be detected in the H$\beta$ and \ion{He}{i} line fluxes. 
The average Balmer decrement, measured as the ratio of H$\alpha$ to H$\beta$ flux, confirms that 
the observed chromospheric emission is mainly coming from extended facular regions.  

The H$\alpha$ and \ion{Ca}{ii} flux curves are well fitted by a model with two circular bright spots (plages), 
also performed with \textsc{Macula}, 
with sizes similar to the underlying spots and a flux contrast of 5 for the H$\alpha$ and about 3 for the 
\ion{Ca}{ii} H\&K and IRT fluxes, as expected from the different modulation amplitude. 
The flux curves are nearly anticorrelated with the photometry, even if the plage at a lower latitude precedes in
longitude, by about 20$\degr$--40$\degr$, the corresponding photospheric spot. 
This finding is not anomalous since lead and lag between plages and spots have been observed in the Sun and are 
frequently invoked as the cause of phase shifts between light curves and chromospheric line fluxes observed both in
young active stars and in RS~CVn systems. 

\begin{acknowledgements}

The authors are grateful to the anonymous referee for a careful reading of the paper and valuable comments. 
We thank the CAHA and OACt teams for the assistance during the observations.
This work has been supported by the Italian {\em Ministero dell'Istruzione, Universit\`a e  Ricerca} (MIUR) and by the 
{\em Regione Sicilia} which are gratefully acknowledged. 
The ESO Scientific Visitor Programme enabled ZsK and KB to visit ESO Garching during the preparation of the paper.
ZsK acknowledges the support of the Hungarian OTKA grants K-68626 and K-81421.
This research has made use of SIMBAD and VIZIER databases, operated at CDS, Strasbourg, France. 

\end{acknowledgements}

\bibliographystyle{elsarticle-harv}

\begin{thebibliography}{}
\bibitem[Ayres et al.(1986)]{Ayres86}
Ayres, T. R., Testerman, L., \& Brault, J. W. 1986, \apj, 304, 542
\bibitem[Barden(1985)]{Bar85}
Barden, S. C. 1985, \apj, 295, 162
\bibitem[Barnes \& Evans(1976)]{Barnes1976}
Barnes, T. G., \& Evans, D. S. 1976, \mnras, 174, 489
\bibitem[Barnes \& Sofia(1996)]{Barnes96}
Barnes, S., \& Sofia, S. 1996, \apj, 462, 746
\bibitem[Biazzo et al.(2007)]{Biazzo07b}
Biazzo, K., Frasca, A., Henry, G. W., Catalano, S., \& Marilli, E. 2007, \apj, 656, 474
\bibitem[Biazzo et al.(2009)]{Biazzo09}
Biazzo, K., Frasca, A., Marilli, E., et al. 2009, \aap, 499, 579
\bibitem[Bus\`a et al.(2007)]{Busa07}
Bus\`a, I., Aznar Cuadrado, R., Terranegra, L., et al. 2007, \aap, 466, 1089
\bibitem[Buzasi(1989)]{Buza89} 
Buzasi, D. L. 1989, Ph.D. Thesis, Pennsylvania State Univ.
\bibitem[Catalano et al.(1998)]{Cata98}  Catalano S., Lanza A.F., Brekke P., Rottmann G.J., \& Hoyng P., 1998,
in  Cool Stars, Stellar Systems and the Sun, ed. R.A Donahue and J.A. Bookbinder, ASP Conf. Ser., 154, 584 
\bibitem[Catalano et al.(2000)]{Cata00}  Catalano S., Rodon\`o M., Cutispoto G., et al., 2000, 
in Variable Stars as Essential Astrophysical Tools, ed. C. $\dot{\rm I}$bano\v{g}lu,  Kluwer Academic Publishers, p.~687 
\bibitem[Chester(1991)]{Chester91}
Chester, M. M. 1991, PhD Thesis, Pennsylvania State Univ.
\bibitem[Cutispoto et al.(2002)]{Cuti02}
Cutispoto, G., Pastori, L., \& Pasquini, L. 2002, \aap, 384, 491 
\bibitem[Cox(2000)]{Cox00} Cox, A. N. 2000, Allen's Astrophysical Quantities (4th ed.), 
New York: AIP Press and Springer-Verlag
\bibitem[Duncan et al.(1991)]{Duncan91}
Duncan D.K., Vaughan A.H., Wilson O.C., et al. 1991, \apjs, 76, 383 
\bibitem[{Ellison(1952)}]{Elli52}
 Ellison, M.~A. 1952, \mnras, 112, 679
\bibitem[Fekel(1997)]{Fekel97}
Fekel, F. C. 1997, \pasp, 109, 514
\bibitem[Flaccomio et al.(2005)]{Flaccomio05}
Flaccomio, E., Micela, G., Sciortino, S., et al. 2005, \apjs, 160, 450
\bibitem[Flower(1996)]{Flower96}
Flower, P. J. 1996, \apj, 469, 355
\bibitem[Frasca \& Catalano(1994)]{Fra94}
Frasca, A., \& Catalano, S. 1994, \aap, 284, 883
\bibitem[Frasca et al.(2000)]{Frasca00}
Frasca, A., Freire Ferrero, R., Marilli, E., \& Catalano, S. 2000, \aap, 364, 179 
\bibitem[Frasca et al.(2005)]{Frasca05}
Frasca, A., Biazzo, K., Catalano, S., et al. 2005, \aap, 432, 647
\bibitem[Frasca et al.(2006)]{Frasca06}
Frasca, A., Guillout, P., Marilli, E., et al. 2006, \aap, 454, 301
\bibitem[Gunn \& Stryker(1983)]{GunnStryker} 
Gunn, J., \& Stryker, L. L. 1983, \apjs, 52, 121
\bibitem[Gunn et al.(1996)]{Gunn1996}
Gunn, A. G., Hall, J. C., Lockwood, G. W., \& Doyle, J. G. 1996, \aap, 305, 146
\bibitem[Hall \&  Ramsey(1992)]{Hall92} 
Hall, J. C., \&  Ramsey, L. W. 1992, \aj, 104, 1942
\bibitem[Harlan(1969)]{Harlan69}
Harlan, E. A. 1969, \aj, 74, 916
\bibitem[Hauschildt et al.(1999)]{Haus99}
Hauschildt, P. H., Allard, F., Ferguson, J., Baron, E., \& Alexander, D. R. 1999, \apj, 525, 871
\bibitem[Herbig(1985)]{Herb85}
Herbig, G. H. 1985, \apj, 289, 269
\bibitem[Huenemoerder(1986)]{Huene86}
Huenemoerder, D. P. 1986, \aj, 92, 673
\bibitem[Huber et al.(2009)]{Huber09}
Huber, K. F., Wolter, U., Czesla, S. et al. 2009, \aap 501, 715
\bibitem[J\"arvinen et al.(2008)]{Jarvi08}
J\"arvinen, S. P., Korhonen, H., Berdyugina, S. V., et al. 2008, \aap, 488, 1047 
\bibitem[Jeffers \& Donati(2008)]{Jeffers08}
Jeffers, S. D., \& Donati, J.-F. 2008, \mnras, 390, 635
\bibitem[Krishnamurthi et al.(1997)]{Krishna97}
Krishnamurthi, A., Pinsonneault, M. H., Barnes, S., \& Sofia, S. 1997, \apj, 480, 303
\bibitem[Kurucz(1993)]{Kuru93}
Kurucz, R. L. 1993, ATLAS9 Stellar Atmosphere Programs and 2 km\,s$^{-1}$ grid, (Kurucz CD-ROM \# 13)   
\bibitem[{LaBonte(1986a)}]{LaBo86a}
 LaBonte B.~J. 1986a, \apjs, 62, 229
\bibitem[{LaBonte(1986b)}]{LaBo86b}
 LaBonte B.~J. 1986b, \apjs, 62, 241
\bibitem[Landman \& Mongillo(1979)]{LandMong79}
Landman, D. A., \& Mongillo, M. 1979, \apj, 230, 581 		   
\bibitem[Landolt(1992)]{Lan92}       
Landolt, A. U. 1992, \aj, 104, 340
\bibitem[Linsky(1970)]{Linsky70}
Linsky, J. L. 1970, \solphys 11, 355
\bibitem[Lo Presti \& Marilli(1993)]{LoPr93}      
Lo Presti, C., \& Marilli, E. 1993,  PHOT -- Photometrical Data Reduction Package. 
Internal report of Catania Astrophysical Observatory N.~2/1993 			    
\bibitem[MacGregor \& Brenner(1991)]{MacGregor91}
MacGregor, K. B., \& Brenner, M. 1991, \apj, 376, 204
\bibitem[Marilli et al.(1997)]{Marilli97}
Marilli, E., Catalano, S., \& Frasca, A. 1997, \memsai, 68, 895
\bibitem[Marsden et al.(2006)]{Marsden06}
Marsden, S. C., Donati, J.-F., Semel, M., Petit, P., \& Carter, B. D. 2006, \mnras, 370, 468
\bibitem[Montes et al.(1995)]{Montes95}
Montes, D., Fern\'andez-Figueroa, M. J., De Castro, E., \& Cornide, M. 1995, \aaps, 109, 135
\bibitem[Montes et al.(2001)]{Montes01}
Montes, D., L\'opez-Santiago, J., G\'alvez, M. C., et al. 2001, \mnras, 328, 45 
\bibitem[{Noyes et al.(1970)}]{Noyes70}
Noyes, R. W., Withbroe, G. L., \& Kirshner, R. P. 1970, \solphys, 11, 388 
\bibitem[Oja(1987)]{Oja87}
Oja T. 1987, \aaps 71, 561
\bibitem[{Oranje(1983)}]{Oranje83}
 Oranje, B.~J. 1983, \aap, 124, 43
\bibitem[Palla \& Stahler(1999)]{palla1999}
Palla, F., \& Stahler, S. W. 1999, \apj 525, 772
\bibitem[Pavlenko \& Magazz\`u(1996)]{PavMag96}
Pavlenko, Y. V., \& Magazz\`u, A. 1996, \aap, 311, 961
\bibitem[Perryman et al.(1997)]{HIPPA97}
Perryman, M. A. C., Lindegren, L., Kovalevsky J., et al. 1997, \aap, 323, L49 
\bibitem[Pfeiffer et al.(1998)]{Pfeiffer1998}
Pfeiffer, M. J., Frank, C., Baumueller, D., et al. 1998, \aaps, 130, 381
\bibitem[Press et al.(1986)]{Press86}    
Press, W. H., Flannery, B. P., Teukolsky, S. A., \& Vetterling, W.T. 1986, Numerical Recipes. 
\bibitem[Prosser et al.(1996)]{Prosser96}    
Prosser, C. F., Randich, S., Stauffer, J. R., Schmitt, J. H. M. M., \& Simon, T. 1996, \aj, 112, 1570
\bibitem[Prugniel \& Soubiran(2001)]{Prugniel01}
Prugniel, P., \& Soubiran, C. 2001, \aap, 369, 1048
\bibitem[Rachford \& Foight(2009)]{Rachford09}
Rachford, B. L., \& Foight, D. R. 2009, \apj 698, 786
\bibitem[Radick et al.(1987)]{Radick87}
Radick, R. R., Thompson, D. T., Lockwood, G. W., Duncan, D. K., \& Baggett, W.E. 1987, \apj, 321, 459
\bibitem[Randich(1997)]{Randich97}
Randich S. 1997, \memsai, 68, 971
\bibitem[Randich et al.(2006)]{Randich2006} Randich, S., Sestito,
P., Primas, F., et al. 2006, \aap, 450, 557
\bibitem[Rice et al.(1989)]{Rice89}
Rice, J. B., Wehlau, W. H., \& Khokhlova, V. L. 1989, \aap, 208, 179
\bibitem[{Roberts et al.(1987)}]{Roberts}
Roberts, D. H., Lehar, J., \& Dreher, J. W., 1987, \aj, 93, 968
\bibitem[{Rossiter(1924)}]{Rossiter1924}
Rossiter, R. A. 1924, \apj, 60, 15
\bibitem[\protect\citeauthoryear{Siess et al.}{2000}]{Siess00}
Siess L., Dufour E., \& Forestini M. 2000, \aap, 385, 593
\bibitem[Simkin(1974)]{Simkin1974}
Simkin, S. M. 1974, \aap, 31, 129
\bibitem[Sneden(1973)]{sneden1973} Sneden, C. 1973, \apj, 184, 839
\bibitem[Soderblom et al.(1993)]{Soderblom1993}			
Soderblom, D. R., Jones, B. F., \& Balachandran, S., et al. 1993, \aj, 106, 1059
\bibitem[Soderblom et al.(1998)]{Soderblom1998}	
Soderblom, D. R., King, J. R., \& Henry, T. J. 1998, \aj, 116, 396
\bibitem[Stassun et al.(2004)]{Stassun04}
Stassun, K. G., Ardila, D. R., Barsony, M., Basri, G., \& Mathieu, R. D. 2004, \aj, 127, 3537
\bibitem[St{\c e}pie\'n et al.(2004)]{Stepien01}
St{\c e}pie\'n, K., Schmitt J. H. M. M., \& Voges, W. 2001, \aap, 370, 157 
\bibitem[Strassmeier et al.(2003)]{Strass03}
Strassmeier, K. G., Pichler, T., Weber, M., \& Granzer, T. 2003, \aap, 411, 595

\end{thebibliography}

\Online

\begin{table}
\caption{$BV$ Johnson photometry of HD~171488.} 
\label{tab:phot} 
\begin{tabular}{l l l l l} 
\hline\hline 
\noalign{\medskip}
 HJD          &   $V$	 &  err    &  $B-V$   &   err   \\
 2\,400\,000+ &  (mag)   &   (mag)  &  (mag)  &  (mag)  \\
\noalign{\medskip}
\hline	     			      	   	 	    	           	  	     	   	   	   	      	   	       
 53962.40200  &  7.429 &  0.006 & 0.614 &  0.004  \\
 53962.40332  &  7.430 &  0.005 & 0.610 &  0.004  \\
 53962.44950  &  7.442 &  0.004 & 0.605 &  0.009  \\
 53962.45024  &  7.430 &  0.006 & 0.615 &  0.008  \\
 53962.51545  &  7.415 &  0.009 & 0.621 &  0.012  \\
 53962.51668  &  7.445 &  0.008 & 0.604 &  0.015  \\
 53963.32720  &  7.435 &  0.005 & 0.606 &  0.010  \\
 53963.32861  &  7.434 &  0.009 & 0.610 &  0.006  \\
 53963.36905  &  7.416 &  0.005 & 0.613 &  0.008  \\
 53963.37295  &  7.431 &  0.009 & 0.605 &  0.011  \\
 53963.40868  &  7.421 &  0.006 & 0.613 &  0.006  \\
 53963.40989  &  7.424 &  0.004 & 0.608 &  0.005  \\
 53963.45009  &  7.435 &  0.008 & 0.607 &  0.014  \\
 53963.45113  &  7.439 &  0.006 & 0.602 &  0.008  \\
 53963.49762  &  7.434 &  0.009 & 0.614 &  0.018  \\
 53963.49901  &  7.435 &  0.012 & 0.600 &  0.019  \\
 53964.31067  &  7.452 &  0.006 & 0.621 &  0.008  \\
 53964.31246  &  7.445 &  0.004 & 0.606 &  0.002  \\
 53964.36066  &  7.436 &  0.006 & 0.611 &  0.009  \\
 53964.36211  &  7.428 &  0.006 & 0.615 &  0.005  \\
 53964.42000  &  7.417 &  0.007 & 0.611 &  0.003  \\
 53964.42103  &  7.438 &  0.004 & 0.609 &  0.008  \\
 53964.46390  &  7.430 &  0.010 & 0.605 &  0.006  \\
 53964.46498  &  7.449 &  0.004 & 0.612 &  0.008  \\
 53964.50762  &  7.416 &  0.013 & 0.615 &  0.019  \\
 53964.50945  &  7.417 &  0.007 & 0.617 &  0.010  \\
 53965.30566  &  7.444 &  0.005 & 0.611 &  0.009  \\
 53965.30695  &  7.456 &  0.008 & 0.615 &  0.009  \\
 53965.33133  &  7.446 &  0.004 & 0.613 &  0.004  \\
 53965.33242  &  7.455 &  0.007 & 0.607 &  0.004  \\
 53965.38597  &  7.454 &  0.005 & 0.604 &  0.006  \\
 53965.38770  &  7.454 &  0.002 & 0.608 &  0.005  \\
 53965.45028  &  7.464 &  0.008 & 0.607 &  0.010  \\
 53965.45120  &  7.453 &  0.005 & 0.619 &  0.003  \\
 53965.49642  &  7.455 &  0.004 & 0.613 &  0.009  \\
 53965.49818  &  7.445 &  0.002 & 0.633 &  0.012  \\
 53966.35221  &  7.446 &  0.008 & 0.605 &  0.006  \\
 53966.35356  &  7.432 &  0.004 & 0.605 &  0.006  \\
 53966.41324  &  7.423 &  0.005 & 0.608 &  0.016  \\
 53966.41460  &  7.428 &  0.005 & 0.612 &  0.008  \\
 53966.46407  &  7.404 &  0.008 & 0.611 &  0.009  \\
 53966.46501  &  7.406 &  0.004 & 0.600 &  0.004  \\
 53967.29996  &  7.383 &  0.016 & 0.605 &  0.016  \\
 53967.30349  &  7.399 &  0.004 & 0.602 &  0.004  \\
 53967.37942  &  7.406 &  0.004 & 0.602 &  0.008  \\
 53967.38090  &  7.405 &  0.006 & 0.607 &  0.005  \\
 53967.44171  &  7.389 &  0.008 & 0.609 &  0.008  \\
 53967.44372  &  7.397 &  0.004 & 0.613 &  0.004  \\
 53967.51569  &  7.428 &  0.014 & 0.624 &  0.011  \\
 53968.31638  &  7.453 &  0.010 & 0.615 &  0.009  \\
 53968.31768  &  7.441 &  0.007 & 0.625 &  0.010  \\
 53968.38402  &  7.429 &  0.005 & 0.619 &  0.005  \\
 53968.38599  &  7.438 &  0.005 & 0.622 &  0.007  \\
 53968.43591  &  7.425 &  0.011 & 0.615 &  0.012  \\
 53968.43727  &  7.426 &  0.007 & 0.614 &  0.014  \\
 53969.30101  &  7.443 &  0.010 & 0.603 &  0.015  \\
 53969.30235  &  7.448 &  0.010 & 0.611 &  0.015  \\
 53969.37936  &  7.460 &  0.010 & 0.610 &  0.007  \\
 53969.38200  &  7.460 &  0.008 & 0.616 &  0.006  \\
 53969.46843  &  7.451 &  0.010 & 0.617 &  0.008  \\
 53969.46947  &  7.475 &  0.003 & 0.615 &  0.005  \\
 53969.52846  &  7.475 &  0.010 & 0.634 &  0.013  \\
 53969.52938  &  7.465 &  0.015 & 0.621 &  0.015  \\
\hline 
\end{tabular}
\end{table}

\end{document}